\documentclass[preprint, 11pt, a4paper]{aastex631}
\usepackage[english]{babel}

\usepackage{amsmath}

\usepackage{graphicx}
\usepackage{natbib}

\newcommand{\kms}{~km~s$^{-1}$}
\newcommand{\obs}{_{\text{obs}}}
\newcommand{\los}{_{\text{LOS}}}
\newcommand{\tot}{_{\text{tot}}}
\newcommand{\E}{_{\text{E}}}
\newcommand{\glow}{_{\text{glow}}}
\newcommand{\lya}{Lyman-$\alpha$~{}}
\newcommand{\vD}{\ensuremath{v_\text{D}}}
\newcommand{\myvec}[1]{\ensuremath{\boldsymbol{{#1}}}}

\begin{document}
\title{WawHelioGlow: a model of the heliospheric backscatter glow. I. Model definition}
\shorttitle{WawHelioGlow}
\shortauthors{Kubiak, Bzowski et al.}

\correspondingauthor{M. Bzowski}
\email{bzowski@cbk.waw.pl}

\author[0000-0002-5204-9645]{M. A. Kubiak}
\affil{Space Research Centre PAS (CBK PAN),\\
Bartycka 18A, 00-716 Warsaw, Poland}

\author[0000-0003-3957-2359]{M. Bzowski}
\affil{Space Research Centre PAS (CBK PAN),\\
Bartycka 18A, 00-716 Warsaw, Poland}

\author[0000-0002-6569-3800]{I. Kowalska-Leszczynska}
\affil{Space Research Centre PAS (CBK PAN),\\
Bartycka 18A, 00-716 Warsaw, Poland} 

\author[0000-0003-3484-2970]{M. Strumik}
\affil{Space Research Centre PAS (CBK PAN),\\
Bartycka 18A, 00-716 Warsaw, Poland} 

\begin{abstract}
The helioglow is a fluorescence of interstellar atoms inside the heliosphere, where they are excited by the solar EUV emission. So far, the helioglow of interstellar H and He have been detected. The helioglow features a characteristic distribution in the sky, which can be used to derive both the properties of interstellar neutral gas and those of the solar wind. This requires a simulation model capable of catching with a sufficient realism the essential coupling relations between the solar factors and interstellar. The solar factors include the solar wind flux and its variation with time and heliolatitude, as well as the heliolatitude and time variation of the solar EUV output. The ISN gas inside the heliosphere features a complex distribution function, which varies with time and location. The paper presents the first version of a WawHelioGlow simulation model for the helioglow flux using an optically thin, single scattering approximation. The helioglow computations are based on a sophisticated kinetic treatment of the distribution functions of interstellar H and He provided by the (n)WTPM model. The model takes into account heliolatitudinal and spectral variations of the solar EUV output from observations. We present a formulation of the model and the treatment of the solar spectral flux. The accompanying Paper II illustrates details of the line of sight evolution of the elements of the model and a brief comparison of results of the WawHelioGlow code with selected sky maps of the hydrogen helioglow, obtained by the SWAN instrument onboard the SOHO mission.

\end{abstract}

\section{Introduction}
\label{sec:intro}
\noindent
The heliospheric backscatter glow is a fluorescence of the interstellar neutral (ISN) gas inside the heliosphere, resonantly excited by the solar EUV radiation. It is an observationally discovered phenomenon \citep{bertaux_blamont:71, thomas_krassa:71} that revealed the existence of the heliosphere \citep{blum_fahr:70b} around the Sun that moves through a warm, magnetized, partially ionized cloud of interstellar matter. 

A helioglow photon is created when a neutral atom is excited by a photon of an appropriate frequency, and subsequently de-excited with emission of a photon with a slightly different frequency in a random direction. Details of this process have been extensively discussed by \citet{brasken_kyrola:98}. At a given location in space, the atoms forming the local population of ISN gas are excited with a rate that depends on one hand on the magnitude of the illuminating spectral flux, and on the other hand on the distribution of radial speeds of the atoms, which are responsible for the Doppler-tuning of individual atoms to the spectral flux. All atoms from the given population that have identical radial velocities have identical probabilities of being excited. Depending on the species, either all atoms are deep within the Doppler width of the illuminating solar line (as is the case for H) and are subject to excitation, or the illuminating line is so narrow that only a portion of the atoms from the given population are within its span, and the rest does not contribute to the formation of the helioglow (as may happen at certain locations for He).

In the atom de-excitation following immediately the excitation, photons are emitted at random directions, given by the so-called scattering function, which depends on the species. Thus, the gas in the given location becomes a volume source of resonant radiation. An observer at a selected vantage point in space measures the intensity of the helioglow (regardless of its wavelength within a certain spectral band), i.e., counts the photons that reach its vantage point from a line extending towards a selected direction from the observer to infinity. Conversely, in a given location in space only the photons that are directed towards the observer will contribute to the helioglow signal that this observer registers. 

The problem of calculating the intensity of the heliospheric backscatter glow of ISN gas has been dealt with by many authors, including \cite{weller_meier:74, meier:77a, keller_thomas:79a, keller_etal:81a, quemerais_bertaux:93a, quemerais:00, quemerais:06a, scherer_fahr:96, fayock_etal:13a}. All these models derive from a series of seminal papers by Hummer \citep{hummer:62a,hummer:64a, hummer:68a, hummer:69a, hummer:69b} and assume that the temperature of the ISN gas inside the heliosphere is isotropic. However, the anisotropy of the distribution function of ISN gas within a few au from the Sun is strong and can be adequately represented neither by an isotropic nor anisotropic Maxwell-Boltzmann function \citep[see, e.g.,][]{fahr:79,kubiak_etal:19a,sokol_etal:15a, sokol_etal:19c} even when one neglects the charge exchange reaction between the charged and neutral interstellar components in the outer heliosheath. 

In this paper, we derive a model of the intensity of the backscatter resonance glow of ISN H and He taking into account the full distribution function returned by the hot model of ISN gas \citep{fahr:78,fahr:79}, calculated using the nWTPM model of ISN gas \citep{tarnopolski_bzowski:09, sokol_etal:15a} where a fully kinetic approach is applied without any analytic approximation for the velocity distribution function inside the heliosphere. We present a first version of the WawHelioGlow model, which uses the optically thin approximation and ignores absorption of the solar illuminating radiation by the ISN gas, as well as effects related to finite optical thickness of the ISN gas. These effects will be included in future versions of the model. In the present version, we implement effects of the time- and heliolatitude-dependent ionization of ISN gas and radiation pressure (which is also dependent on the atom radial velocity), as well as the currently most reliable estimates for the evolution of the illuminating solar spectral lines of the \lya and He I 58.4 nm radiation. 

The model is devised as a tool in analysis of future observations from the GLOWS experiment on the NASA Interstellar Mapping and Acceleration Probe mission \citep[IMAP][]{mccomas_etal:18a}. The objective of the paper is to present the baseline version of the model. A qualitative comparison of the model results with selected helioglow observations from the Solar Wind ANisotropy experiment onboard the SOlar and Heliospheric Observatory \citep[SWAN/SOHO; ][]{bertaux_etal:95} is presented in the accompanying Paper II \footnote{M.A. Kubiak, WawHelioGlow: a model of the heliospheric backscatter glow. II. The helioglow buildup and potential significance of the anisotropy in the solar EUV output }. An in-depth analysis of these observations is postponed to a future work.

In Section~\ref{sec:modelDefinition} we present the definition of the helioglow intensity for a selected line of sight, differentiating between the cases of the ISN hydrogen and helium glows. In particular, we discuss the models of the solar illumination of ISN H and He by the relevant portions of the solar EUV spectrum. We start with a presentation of the geometry of the line of sight and proceed to discuss the definition of the source function of the helioglow, the scattering phase function, the distribution function of the ISN gas and its partial density, the excitation function of the gas, and the illumination function. We close this long section with presentation of some numerical aspects of the integration of the helioglow intensity, and close the paper with a summary and an outlook leading to the accompanying Paper II. In Appendix \ref{sec:appendixCrossSection} we present some textbook derivations of the photon absorption cross sections and the radiation pressure for ISN H atoms, 
In Appendix~\ref{sec:appendix1}, we summarize available measurements of spectral features of the solar He I 58.4 nm line, to better justify the choices of the parameter values of the model of this line used in the paper.
\section{Model definition}
\label{sec:modelDefinition}

\begin{deluxetable}{lll}
\tablecaption{{\label{tab:nrDef}} The most important quantities defined in the model, with their respective equations}
\tablehead{\colhead{Defined quantity} & \colhead{Equation No} & \colhead{Description}}
\startdata
$ I\glow(\tilde{S}_\sun, \myvec{r}\obs, \hat{\myvec{l}}) $   & \ref{eq:helioglowIntensDef1}, \ref{eq:helioGlowIntensityDef}, \ref{eq:helioGlowIntensityH} & helioglow intensity\\
$ J\glow(\tilde{S}_\sun, \myvec{l})$ & \ref{eq:helioglowSourceFunDef} & helioglow source function \\
$\hat{\myvec{l}}$ & \ref{eq:LineofSight}, Fig.\ref{fig:LOSgeometry} & look (i.e., line of sight) direction \\
$\myvec{r}\los\left(\myvec{l} \right)$ & \ref{eq:rLOS}, Fig.\ref{fig:LOSgeometry} & heliocentric radius-vector of a point belonging to the line of sight (LOS) \\
$ \beta $ & \ref{eq:betaDefinition}, \ref{eq:cosBetaDefinition}, Fig.\ref{fig:LOSgeometry} & scattering phase angle\\
$ S(\myvec{l}, v_r) $ & \ref{eq:defSrcFun} & spectral source function of the helioglow\\
$\psi$  & \ref{eq:defPhaseFunH}, \ref{eq:defPhaseFunHe}  & scattering phase functions \\
$ \eta(\myvec{l}, v _r) $& \ref{eq:partialDens} & partial density of ISN gas, dependent on $v_r$ \\
$\sigma(\nu)$ & \ref{eq:absCrossctnDef2}  & cross section for absorption of a photon by a H, He atom\\
$ E(\tilde{S}_\sun ,\myvec{l}, \nu) , E(\tilde{S}_\sun ,\myvec{l}, v_r) $& \ref{eq:excitationFunDef} , \ref{eq:excitFunWorkDef}, \ref{eq:excitFunWorkDef2} & excitation function of ISN gas\\
$\tilde{S}_\sun(v_r,t) $ & \ref{eq:solSpectrFluxDistance}, \ref{eq:solSpectrFluxDef}, \ref{eq:sSunHeDef} &  illuminating solar spectral flux \\
$\mu$, $\langle \mu \rangle$ & \ref{eq:solarProfileVsDistanceH}, \ref{eq:averageMu} & radiation pressure factors: instantaneous and average \\
$s(v_r) $ & \ref{eq:solarProfileVsDistanceH}, \ref{eq:defHe584LineProfile}, \ref{eq:defKappaProfile}  & model of the solar spectral line profile, unnormalized \\
$ p $ & \ref{eq:normFactHDefinition}, \ref{eq:normFactHeDefinition} & normalization factor for the line profile $s$\\
$ v_D $ & \ref{eq:defineDopplerShift}, \ref{eq:gaussFWHMVsVd}, \ref{eq:gaussFWHMVsVdBkg}, \ref{eq:defKappaFWHM} & Doppler speed \\
$ I_{\sun}(\myvec{r}\los,t) $ & \ref{eq:iLambdaPhiNormalization}, \ref{eq:iSunPhiDefinition}, \ref{eq:iTotVsISun}, \ref{eq:IsunDefinition} & total solar illuminating flux at $\myvec{r}\los$ \\
$ I_{\sun,\text{E}}(\phi,t) $ & \ref{eq:IsunDefinition} & total solar illuminating flux at the reference distance $r\E$ and heliolatitude $\phi$ \\
$ I_{\lambda \phi}(\phi) $    & \ref{eq:iSunPhiDefinition} & solar flux at heliolatitude $\phi$ relative to that at $\phi = 0$  \\
\enddata
\end{deluxetable}

\noindent
The objective of the WawHelioGlow model is to calculate a wavelength-integrated flux density (i.e., the total intensity per steradian) of the heliospheric backscatter glow (the helioglow), observed by an observer at a location given by radius-vector $\myvec{r}\obs$, looking into a direction specified by the spherical coordinates $\lambda\los, \phi\los$. The gas is illuminated by a solar emission line: \lya for H and the 58.4 nm line for He. The spectral intensity of the solar illumination at a given moment $t$ is given by the illuminating solar spectral flux $\tilde{S}_\sun(\nu)$ (Equation  \ref{eq:solSpectrFluxDef}), which varies with time, possibly heliolatitude, and depends on photon frequency $\nu$. It is assumed that photons from the Sun propagate instantaneously and are scattered only once, which greatly simplifies the radiation transfer equation \citep{hummer:69b}. The helioglow intensity $I\glow(\tilde{S}_\sun, \myvec{r}\obs, \hat{\myvec{l}})$ is given by a line-of-sight (LOS) integral of the helioglow source function $J\glow(\tilde{S}_\sun,\myvec{l})$:
\begin{equation}
I\glow(\tilde{S}_\sun, \myvec{r}\obs, \hat{\myvec{l}}) = 
 \int\limits_{0}^l J\glow(\tilde{S}_\sun, \myvec{l})\, d l',
\label{eq:helioglowIntensDef1}
\end{equation} 
where $\myvec{l} = l'\, \hat{\myvec{l}}$. A precise definition of the geometry of the line of sight is given in Section~\ref{sec:LOSDef}. The partial helioglow intensity $I\glow(l)$ is defined so that the upper boundary of the integration is finite and equal to $l$. For $l \rightarrow \infty$, the partial helioglow intensity becomes the full intensity. The helioglow source function $J\glow$ is defined as an integral over atom velocity of the helioglow spectral source function $S(\tilde{S}_\sun, \myvec{l}, v_r)$:
\begin{equation}
J\glow(\tilde{S}_\sun, \myvec{l}) = \int\limits_{-\infty}^{\infty} S(\tilde{S}_\sun(v_r), \myvec{l}, v_r) d {v_r},
\label{eq:helioglowSourceFunDef}
\end{equation}
where we have converted $\tilde{S}_\sun$ from the dependence on photon frequency $\nu$ to that on the radial velocity of atoms $v_r$ due to the Doppler effect (Equation \ref{eq:freqVrDefinition}).  The source function $S$ (Equation \ref{eq:defSrcFun}) is a product of the excitation function $E$ of individual atom (Equation \ref{eq:excitationFunDef}) for a given radial velocity $v_r$, the number of atoms suitable for excitation at this radial speed $v_r$, given by the partial density $\eta$ (Equation \ref{eq:partialDens}), and the probability of emission of the re-emitted photons towards the observer, given by the scattering phase function $\psi$ (Equations \ref{eq:defPhaseFunH}, \ref{eq:defPhaseFunHe}). The atom excitation function $E$ depends on the solar spectral flux $\tilde{S}_\sun$ at a given location in space (Equations \ref{eq:solSpectrFluxDistance}, \ref{eq:solSpectrFluxDef}, \ref{eq:sSunHeDef}) and on the cross section $\sigma$ for photon absorption by an atom of a given species  (Equation \ref{eq:absCrossctnDef2}).

In the following, we define and discuss all relevant factors, starting from the definition of the geometry of the line of sight and proceeding to the aforementioned factors composing the spectral source function. For readers convenience, we summarize in Table~\ref{tab:nrDef} the most important quantities and the equations where they are defined. The geometry of observations assumed in the model is presented in Figure~\ref{fig:LOSgeometry}. 

\subsection{Line of sight $\myvec{l}$ and other geometrical definitions}
\label{sec:LOSDef}
\begin{figure}[ht!]
\centering
\includegraphics[width=0.49\textwidth]{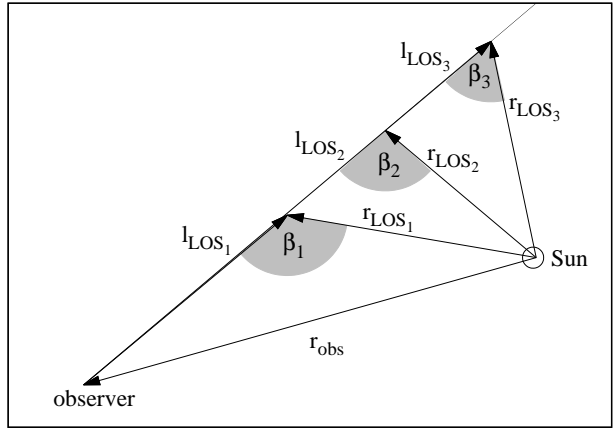}
\caption{The geometry of the line of sight defined in Equations \ref{eq:LineofSight} and \ref{eq:rLOS}, with three example points $\myvec{r}_{\los,{1-3}}$  along this line. The observer is in the location in space defined by the radius-vector $\myvec{r}\obs$, anchored at the Sun. Along the line of sight, the three scattering phase angles $\beta_{1-3}$ (Equations \ref{eq:betaDefinition}, \ref{eq:cosBetaDefinition}) at the three selected points are presented, along with the corresponding lengths along the line of sight $l_{\los,{1-3}}$ (Equation \ref{eq:LineofSight}).}  
\label{fig:LOSgeometry}
\end{figure}
\noindent
The geometry of the ine of sight is presented in Figure~\ref{fig:LOSgeometry}.
The observer position is defined by the vector $\myvec{r}\obs$ in a Sun-centered frame. The line of sight in space is defined by its origin (i.e., $\myvec{r}\obs$) and the look direction $\hat{\myvec{l}}$.
The unit vector $\hat{\myvec{l}}$ is defined by the spherical coordinates $\lambda\los, \phi\los$ of the direction of observation as follows:
\begin{equation}
\myvec{\hat{l}}\left(\lambda\los, \phi\los \right) = 
\left(\begin{array}{c}
\cos \lambda\los\, \cos \phi\los \\
\sin \lambda\los\, \cos \phi\los \\
\sin \phi\los
\end{array}\right),
\label{eq:LineofSight}
\end{equation}
and the points in the 3D space belonging to the line of sight are given by the relation:
\begin{equation}
\myvec{r}\los\left(\myvec{l} \right) = \myvec{r}\obs  \myvec{l} = \myvec{r}\obs  l\, \hat{\myvec{l}}, \text{ for } l \in (0, ..., \infty). 
\label{eq:rLOS}
\end{equation}
Note that a given point belonging to a line of sight can be alternatively and fully equivalently identified either by $\myvec{r}\los$ or $(\myvec{r}\obs,  \myvec{l})$.

For any point $\myvec{r}\los$, the scattering phase angle $\beta$, i.e., the angle at which the radiation incoming at $\myvec{r}\los$ from the Sun is scattered towards the observer at $\myvec{r}\obs$ is defined as the angle between $\myvec{r}\los$ and $-\myvec{\hat{l}}$:
\begin{equation}
\beta \equiv \angle(\myvec{r}\los, -\myvec{\hat{l}}); 
\label{eq:betaDefinition}
\end{equation}
Cosine of this angle is given by the projection of the unit vector $\hat{\myvec{r}}\los$ on the direction towards the observer:
\begin{equation}
\cos \beta = -\hat{\myvec{l}}\cdot \hat{\myvec{r}}\los. 
\label{eq:cosBetaDefinition}
\end{equation}

\subsection{Definition of the spectral source function $S$}
\label{sec:sourceFunctionDef}
\noindent
The spectral source function of the helioglow $S$ is a product of the scattering phase function, partial density of the ISN gas, and the excitation function of this gas. Both the partial density and the excitation function depend on the illuminating solar spectral flux $\tilde{S}_\sun$. The dependence of $E$ on $\tilde{S}_\sun$ is direct and that of the partial density is indirect. The solar spectral flux excites the atoms (the direct dependence), which emit the helioglow, and affects the partial density because it is responsible for the radiation pressure effect, which modifies the trajectories of the atoms in space and thus the distribution function of ISN atoms along the line of sight (the indirect dependence).

The helioglow spectral source function $S(\tilde{S}_\sun, \myvec{l}, v_r)$ at a location $\myvec{l}$ within the line of sight for an ISN atom traveling at a velocity $\myvec{v} = (v_r, v_{T_1}, v_{T_2})$ is defined as follows:
\begin{equation}
S(\tilde{S}_\sun, \myvec{l}, v_r) = 
\psi(\beta)\,
\eta(\myvec{l}, v_r)\,
E(\tilde{S}_\sun, \myvec{l}, v_r).
\label{eq:defSrcFun}
\end{equation}
Here, $v_r = \myvec{v}\cdot \hat{\myvec{r}}\los(\myvec{l})$ is the radial speed of an individual atom from the ISN gas at $\myvec{l}$. The quantities $v_{T_1}, v_{T_2}$ are the  two components of the Cartesian vector perpendicular to the solar direction. In this formulation, the reference system can be arbitrarily selected provided that the $v_r$ axis is heliocentric. Here, we integrate over $v_{T1}, v_{T2}$ to form the partial density $\eta$ of ISN gas. For definition of $\eta$, see Equation \ref{eq:partialDens}. 

$E(\tilde{S}_\sun,\myvec{l}, v_r)$ is the excitation function for individual atoms, and $\psi(\beta)$ is the phase function of the scattering between the direction of incidence given by $\hat{\myvec{r}}\los$ and the direction of re-emission, given by $-\hat{\myvec{l}}$ (Equations \ref{eq:betaDefinition}, \ref{eq:cosBetaDefinition}). Thus, the product $\eta\, E$ gives the number of excited atoms per unit time and unit volume, and $\psi$ determines the fraction of photons resulting from de-excitation that are directed towards the detector. A precise definition of the function $E$ is presented in Section~\ref{sec:excitationFunctionDef}. The partial density $\eta$ is presented in Section \ref{sec:ISNDistrFun}, and the scattering phase function $\psi$ in Section~\ref{sec:phaseFunction}.

\subsection{The phase function $\psi$ }
\label{sec:phaseFunction}
\begin{figure}[ht!]
\centering
\includegraphics[width=0.5\textwidth]{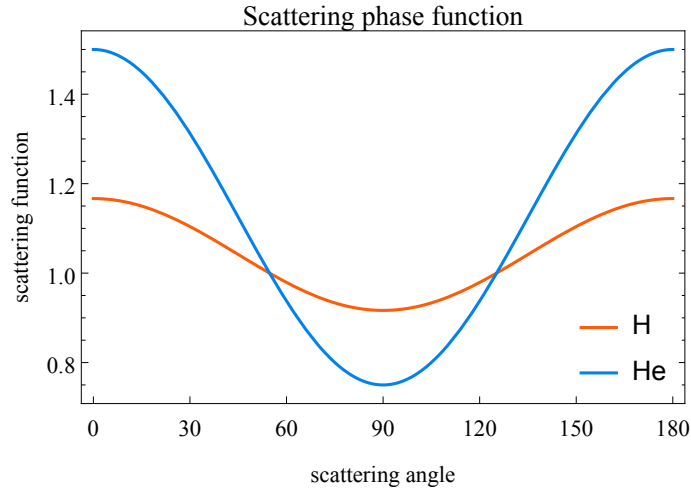}
\caption{Phase function for scattering the solar radiation by ISN H and He atoms, adopted after \citet{meier:77a} and defined in Equations \ref{eq:defPhaseFunH} and \ref{eq:defPhaseFunHe}.}
	\label{fig:phaseFun}
\end{figure}
\noindent
The phase function $\psi(\beta)$ determines the probability that if a photon arrives from the direction $\hat{\myvec{r}}$, it will be re-emitted towards the direction defined by $-\hat{\myvec{l}}$. The angle between these directions is denoted $\beta$, see Section \ref{sec:LOSDef} and Fig.\ref{fig:LOSgeometry}. 
The formulae for the phase functions for H and He, adopted after \citet{meier:77a}, is defined in the following equations: 
\begin{equation}
\psi_{\text{H}}(\beta) = \frac{1}{12}\left(11 + 3 \cos^2 \beta  \right)\, \, \text{for H}, 
\label{eq:defPhaseFunH}
\end{equation} 
\begin{equation}
\psi_{\text{He}}(\beta) =  \frac{3}{4}\left(1 + \cos^2\beta\right)\, \, \text{for He}
\label{eq:defPhaseFunHe}
\end{equation}
and presented in Figure~\ref{fig:phaseFun}.

\subsection{The distribution function of the ISN gas and the partial density $\eta$}
\label{sec:ISNDistrFun}
\noindent
The term $\eta(\myvec{l},v_r)$ in Equation~\ref{eq:defSrcFun} is the partial density of ISN gas at $\myvec{l}$, defined as follows:
\begin{equation}
\eta(\myvec{l}, v _r) = \int \limits_{-\infty}^\infty \int \limits_{-\infty}^\infty f_\text{ISN}(\myvec{l}, v_r, v_{T_1}, v_{T_2})\, d v_{T_1}\, dv_{T_2}.
\label{eq:partialDens}
\end{equation}
In this definition, $f_\text{ISN}(\myvec{l}, \myvec{v})$ is the distribution function of ISN gas at the location $\myvec{l}$ for an atom with a velocity $\myvec{v} = (v_r, v_{T_1}, v_{T_2})$. The magnitude of $f_\text{ISN}(\myvec{l}, \myvec{v})$ depends on the survival probability of ISN atoms against ionization losses and on the radiation pressure, calculated for this time and location. Radiation pressure is conveniently expressed by a coefficient of compensation of solar gravity by the radiation pressure force $\mu$. In the heliosphere, only H is affected by this effect; for He, we adopt $\mu \equiv 0$. The survival probability is determined by a function $\beta_I(\phi, t)$ that calculates the instantaneous ionization rate of a given species for a given heliolatitude $\phi$ and time $t$. The functions $\beta_I$ and $\mu$ are used in the calculation of the distribution function $f_\text{ISN}$, which involves tracking the atom trajectories in time and space between $\myvec{r}\los$ and the entrance to the heliosphere \citep{rucinski_bzowski:95b, rucinski_etal:03, bzowski_etal:02}. Therefore, formally $f_\text{ISN}$ is a function of these quantities as well as the velocity and $\myvec{r}\los$: $f_\text{ISN}\equiv f_\text{ISN}(\myvec{r}\los, \myvec{v}\los, \beta_I, \mu)$, but further in the paper we will omit them for brevity. 

The distribution function is marginalized over the two components $v_{T_1}, v_{T_2}$ perpendicular to $v_r$ and to each other; integrating $\eta(\myvec{l},v_r)$ over $v_r$ yields $n(\myvec{l})$ -- the local density of ISN gas.  

Basically, the distribution function can be evaluated using any reasonable approximation, including that of the hot model of the ISN gas \citep[e.g.][]{fahr:78, wu_judge:79a} or in a simplified form that of the Maxwell-Boltzmann function with the density, velocity, and temperature obtained from moments of the function derived from the hot model. In the latter case, some of the integrals in the derivation of the glow intensity can be performed analytically.

In this paper, the distribution function of ISN gas at $\myvec{l}$ is calculated fully kinetically, using the numeric version of the Warsaw Test Particle Model (nWTPM) of the ISN gas. For H, it is computed using a version of this model in which radiation pressure is taken into account in addition to solar gravity \citep{tarnopolski_bzowski:09}. The model of radiation pressure is based precisely on the same profile of the solar \lya radiation as that used for the illuminating function  (Section~\ref{sec:totalIntensity}). For He, the density is computed using the version presented by \citet{sokol_etal:15b}, where the motion of He atoms is governed solely by solar gravity. 

The WTPM method of calculation of the local distribution function of ISN gas was originally developed by \citet{rucinski_bzowski:95b}. The local distribution function is calculated using the hot-model paradigm \citep{fahr:78, fahr:79, wu_judge:79a}, but the atom trajectories are tracked numerically with the effects of time-, latitude- and radial velocity-dependent radiation pressure taken into account \citep{tarnopolski_bzowski:09}. The survival probabilities of the atoms against ionization processes, which also vary with the solar distance, heliolatitude, and time \citep{bzowski_etal:12b, bzowski_etal:13b}, are calculated by numerical integration of the ionization probability along the trajectories \citep{bzowski_etal:02}. In this paper, for illustration purposes, we adopt a model of these factors by \citet{sokol_etal:20a}, but any reasonable ionization model can be used instead. In particular, what counts is the calculation of the ionization losses of ISN atoms, i.e., the total rate of their ionization, without the need to calculate the rates of the constituent ionization processes separately. This feature enables determining the spatial structure of the ionization rates of ISN gas based on observations of the helioglow, as it has been successfully performed in the past \citep[see, e.g.][]{bertaux_etal:99, bzowski:03, bzowski_etal:03a, katushkina_etal:19a, koutroumpa_etal:19a, kyrola_etal:98, lallement_etal:10b, summanen_etal:93, summanen_etal:97, summanen:00}.

Since the helioglow model is optically thin, single scattering, it is possible to account for multiple populations of ISN gas by calculating their local distribution functions and the helioglow contributions separately and co-add the results. For H, we take the primary and secondary populations of ISN gas with the inflow parameters obtained by \citet{bzowski_etal:08a} from analysis of pickup ion observations on Ulysses, with subsequent adjustment of their parameters by \citet{IKL:18b}. This latter adjustment was to account for a larger temperature of the primary population, obtained by \citet{bzowski_etal:15a} from observations of ISN He on IBEX, and for a distortion of the heliosphere from axial symmetry by the interstellar magnetic field, which results in an angular shift of the apparent inflow direction of the secondary population from that of the primary \citep{kubiak_etal:16a}.

Specifically, we adopted for the primary population of hydrogen the J2000 longitude 255.745\degr, latitude 5.169\degr, velocity 25.784 \kms, temperature 7443 K, density 0.031 cm$^{-3}$. For the secondary population, we adopted the longitude 251.57\degr, the latitude 11.95\degr, the velocity 18 \kms, the temperature $16\,300$ K, and the density 0.054189 cm$^{-3}$. For He, we adopted one population, with the parameters identical to those for the primary H population with the density equal to 0.015 cm$^{-3}$ \citep{gloeckler_etal:04b, mobius_etal:04a}; the secondary population was neglected. 

In the optically thin approximation, the absolute intensity of the helioglow is linearly proportional to the absolute density of the ISN gas in front of the heliosphere. This approximation will not exactly hold when multiple scattering effects are taken into account.

It is also possible to use an alternative method to calculate the distribution function of ISN He, defined by \citet{bzowski_etal:17a} \citep[see also][]{bzowski_etal:19a}. In this method, the primary and the secondary populations are not clearly separated. The local distribution function is obtained within the hot-model paradigm extended by solutions of the production and loss equations of ISN He atoms along individual atom trajectories from the unperturbed interstellar medium through the perturbed interstellar plasma outside the heliopause down to a given location at $\myvec{r}\los$. This approach, currently available for ISN He, may be adopted in a future version of the model.

The distribution function of ISN gas  $f_{\text{ISN}}(\myvec{l}, \myvec{v})$ bears imprints from the ionization processes and radiation pressure, which vary with time. This variation adds to the time variation of the spectral source function $S$ resulting from variations in the solar spectral illumination flux $\tilde{S}_\sun$, but while the time scale of the latter is on the order of the solar rotation period, the time scale of the former is more typically from a month to a year, depending on the distance from the Sun. 

\subsection{The excitation function $E$}
\label{sec:excitationFunctionDef}
\noindent
The excitation function $E(\tilde{S}_\sun,\myvec{l}, v_r)$ is the rate (a frequency) of exciting an atom traveling with a radial speed $v_r$ from energy level 1 to level 2 given the illuminating spectral flux $\tilde{S}_\sun(v_r)$. The atom can be excited by photons from a narrow spectral range around the central frequency $\nu_0$, corresponding to the energy difference $h\,\nu_0$ between the two energy levels ($h$ being the Planck constant). The cross section for absorption of a photon in the rest frame of the atom is given by the following formula (see Equation \ref{eq:intCrossSecF1}, derived in Appendix \ref{sec:crossSectn}): 
\begin{equation}
\sigma_{\nu} = \frac{\pi\, e^2}{m_e\,c}\, f_\text{osc}.
\label{eq:absCrossctnDef2}
\end{equation}

When an atom is traveling with a velocity $\myvec{v} = (v_r, v_{T_1}, v_{T_2})$ relative to the Sun, then in the atom rest frame the Sun is moving with respect to the atom with the velocity $-(v_r, v_{T_1}, v_{T_2})$ and photons emitted by the Sun are Doppler-shifted by a frequency corresponding to $-v_r$, based on the relation
\begin{equation}
v_r = \left(\frac{\nu}{\nu_0} - 1\right)c = \left(\frac{\lambda_0}{\lambda} -1 \right)\,c. 
\label{eq:freqVrDefinition}
\end{equation} 

The excitation function in the atom rest frame is defined as
\begin{equation}
E(\tilde{S}_\sun(\nu_0, r\los), \myvec{l},\nu) = \int \limits_{0}^{\infty} \tilde{S}_\sun(\nu, r\los) \sigma(\nu_0, \nu) \, d\nu,
\label{eq:excitationFunDef}
\end{equation}
where  
$\tilde{S}_\sun(\nu, r\los)$ 
is a Doppler-shifted illuminating solar spectral flux for the frequency $\nu$ at the atom location in the \lya or 58.4 nm line. Inserting Equation \ref{eq:absCrossctnDef2} and integrating over $\nu$ one obtains the excitation function in the atom rest frame, which can be immediately transferred to the solar inertial frame, yielding
\begin{equation}
E(\tilde{S}_\sun(\nu_0, r\los), \myvec{l}, \nu_0) =  \frac{\pi\, e^2}{m_e\,c}\,f_{\text{osc}}\, \tilde{S}_\sun(\nu_0, r\los),
\label{eq:excitFunWorkDef}
\end{equation}
Converting now the excitation function so that it depends on radial velocity instead of the frequency (Equation \ref{eq:freqVrDefinition}) and assuming there is no absorption of the solar spectral flux between the solar surface and $\myvec{r}\los$ 
\begin{equation}
\tilde{S}_\sun(v_r,r\los) = \tilde{S}_\sun(v_r,r\E) \left(\frac{r\E}{r\los}\right)^2
\label{eq:solSpectrFluxDistance}
\end{equation}
we obtain
\begin{equation}
E(\tilde{S}_\sun(v_r), \myvec{l}, v_r) = \frac{\pi\, e^2}{m_e\,c}\, f_{\text{osc}} \, \tilde{S}_\sun(v_r,r\E) \left(\frac{r\E}{r\los}\right)^2,
\label{eq:excitFunWorkDef2}
\end{equation}
where $r\E$ is the solar distance for which the illuminating solar spectral flux $\tilde{S}_\sun(v_r)$ is known (typically $r\E = 1$~au). As it is implied by this equation, it is assumed that there is no modification of the illuminating solar spectral flux with the distance from the Sun other than scaling  by the square of solar distance. This assumption may be removed in the future. 

\subsection{The illuminating solar spectral flux $\tilde{S}_\sun$}
\label{sec:solarFlux}
\noindent 
The factors relevant for the helioglow production are the spectral shape of the solar EUV emission in spectral bands characteristic for the helioglow of ISN H and He, and the total intensity of radiation within these wavebands in the geometric locations along the line of sight. It is assumed that the illuminating solar spectral flux $\tilde{S}_\sun$ factorizes as follows:
\begin{equation}
\tilde{S}_\sun(v_r) = p\, s(I_{\sun,\text{E}}, v_r).
\label{eq:solSpectrFluxDef}
\end{equation}
The spectral shape of the function $\tilde{S}_\sun$ is defined by a profile function $s(I_{\sun,\text{E}}, v_r)$, which may be unnormalized; normalization is obtained by multiplication by a scaling factor $p$. The factor $p$ may or may not be a function of the solar flux $I_{\sun,\text{E}}$, depending on the definition of $s$. Thus, the profile function $s$ may vary along the line of sight $\myvec{l}$ because $\myvec{l}$ traverses various heliolatitudes and $I_{\sun,E}$ may depend on heliolatitude (Equation \ref{eq:IsunDefinition}). It is assumed that time variation in the solar illuminating flux propagates instantaneously to all points within the LOS. 

The elements of the solar illuminating flux function $I_{\sun,\text{E}}$ and its components are discussed in Sections \ref{sec:iLambdaPhi} and \ref{sec:totalIntensity}, after presentation of the functions $s$ and the factors $p$ for H and He (Sections \ref{sec:solarHLineProfile} and \ref{sec:solarHeLineProfile}). 

\subsubsection{The spectral profile for hydrogen $s_\text{H}$ and the scaling factor $p_\text{H}$}
\label{sec:solarHLineProfile}
\begin{figure*}[ht!]
\centering
\includegraphics[width=0.45\linewidth]{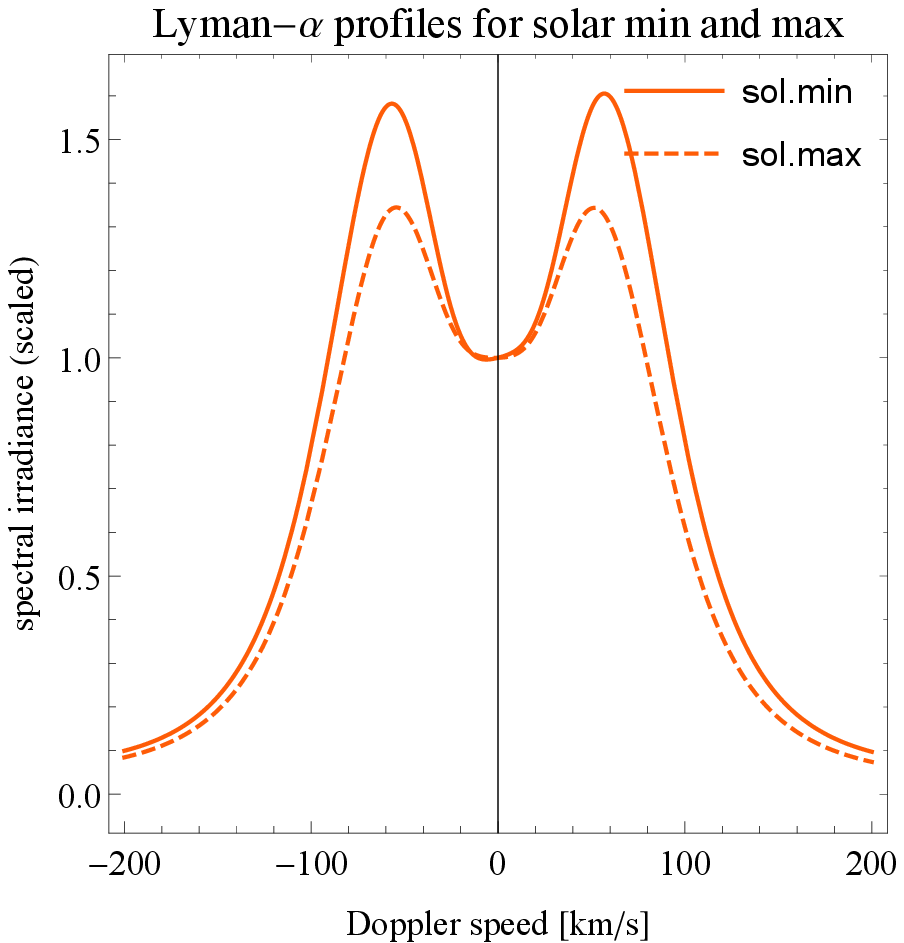}
\includegraphics[width=0.45\linewidth]{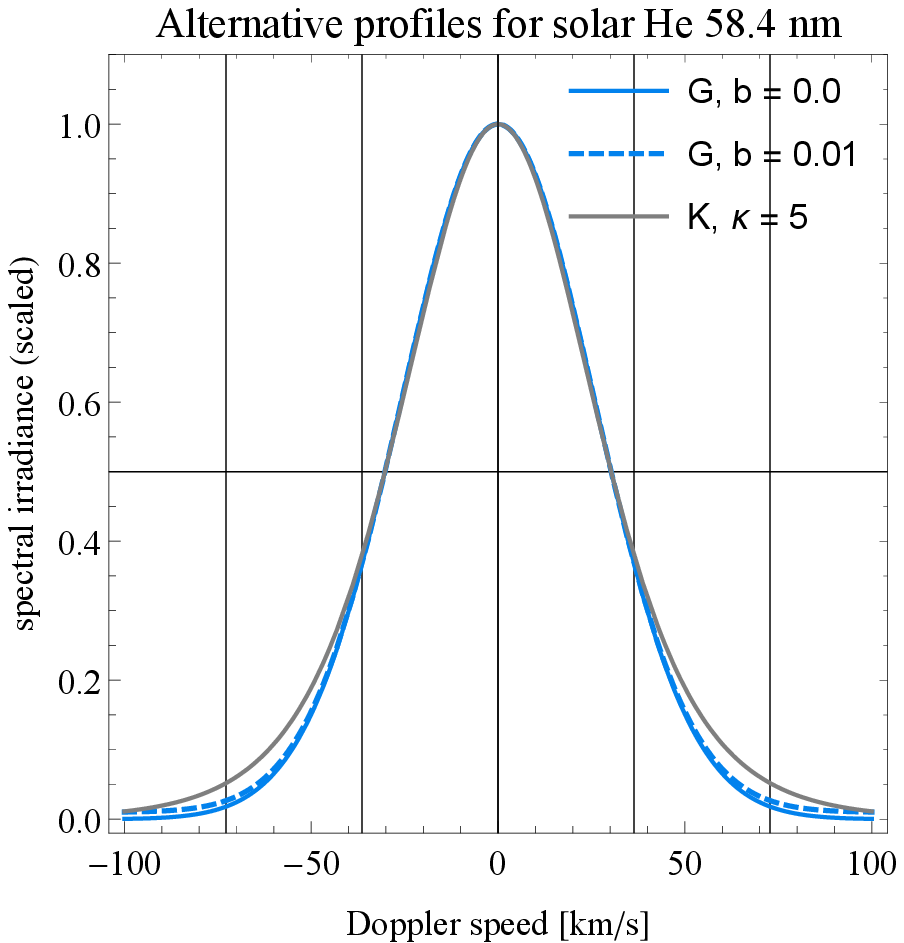}
\caption{Profiles of the solar \lya (left) and He I 58.4 nm lines (right), for illustration purposes normalized to the spectral irradiance at their respective central wavelengths of $\sim  121.6$ nm and $\sim 58.4$ nm. The profiles are defined as a function of the Doppler shifts, expressed in \kms. The left panel presents the shapes of the \lya profile defined by \citet{IKL:20a}, for the solar minimum and maximum conditions, the epochs 1996.43 and 2001.43, respectively. The spectral irradiances for the line center for these epochs are equal to $2.76\times 10^{12}$ and $4.49 \times 10^{12}$~cm$^{-2}$~s$^{-1}$~nm$^{-1}$. This panel illustrates the variation of the profile shape with the solar activity level. The right panel presents alternative profiles for the He I 58.4 nm line, defined in Equations~\ref{eq:defHe584LineProfile} and \ref{eq:gaussFWHMVsVdBkg} for Gaussian profiles (G) with background equal to 0 and 0.01 and for a kappa profile (K) with an identical FWHM, $\kappa = 5$ (Equations \ref{eq:defKappaProfile} and \ref{eq:defKappaFWHM}). The vertical bars mark multiplicities of the adopted $v_{\mathrm{D}}$ speeds. The horizontal bar marks half of the height, to illustrate that with the adopted definitions, the spectral flux defined by the Gaussian and kappa profiles within the FWHM range are practically identical.}
\label{fig:lineProfiles}
\end{figure*}

\noindent
The central wavelength of the \lya line is equal to $\lambda_\text{H} = 121.56701$~nm, and the oscillator strength is $f_{\text{osc},\text{H}} = 0.41641$ \citep{wiese_fuhr:09a}.
The solar \lya line is relatively wide compared with the Doppler range corresponding to the velocity spread of interstellar neutral hydrogen in the heliosphere. The line can be approximately described by a kappa profile sitting on top of a sloped background, with a chromospheric Gaussian-like self-reversal, which results in a characteristic two-horn profile (Figure~\ref{fig:lineProfiles}, left panel). The shape of the profile (e.g., the horn-to-center ratio) evolves with the evolution of the solar activity, as evidenced by the different shapes for the two selected epochs shown in the aforementioned figure. The spectral irradiance is expressed as a function of Doppler speed $v_r$.

The shape of the solar \lya profile and its evolution during the solar activity cycle are known relatively well owing to a series of spectral observations performed by SUMER/SOHO in the years 1996--2009 \citep{lemaire_etal:15a}. Based on these observations, a model of the evolution of solar resonant radiation pressure for H atoms during the cycle of solar activity was developed by \citet{IKL:18a, IKL:18b}, which is parameterized by the composite \lya flux $I_{\text{tot,H}}(t)$ \citep{woods_etal:96a, woods_etal:00}. Here, we adopt the most recent version of this model by \citet{IKL:20a}, which uses the most recent version of $I_{\text{tot,H}}(t)$, compiled by \citet{machol_etal:19a}. Specifically, we use Equation 14 from \citet{IKL:18a} with the parameters from Table 1 in \citet{IKL:20a} and obtain the radiation pressure factor $\mu(I_{\sun,\text{E}}, v_r)$ (Equation \ref{eq:solarProfileVsDistanceH}), i.e., the factor of compensation of the solar gravity force for a selected Doppler velocity. $I_{\sun,\text{E}}(t,\phi)$ is obtained from Equation \ref{eq:IsunDefinition}. In our paper, we apply this result directly in the calculation of the distribution function of ISN H, and after an appropriate rescaling, in the calculation of the solar illumination. 

To this latter end, renormalization of the system defined by \citet{IKL:18a}  must satisfy that in Equation \ref{eq:solarProfileVsDistanceH} we insert $ I_\sun=I_\text{tot,H}$ and  integrate $s_\text{H}( r\E,I_\text{tot,H}(t), v_r)$ over $v_r$ we obtain $I_\text{tot,H}(t)$. The integration interval is defined so that the radial speed boundaries correspond to the wavelength range $(121, 122)$~nm, because this is the interval used when measuring the solar composite \lya flux \citep{woods_etal:05a}. Therefore, the adopted model of the solar \lya line does not require any further normalization other than rescaling to the units of ph cm$^{-2}$ s$^{-1}$ nm$^{-1}$. 
\begin{eqnarray}
s_\text{H}( r\E, I_\sun, v_r)  =  \mu(I_{\sun,\text{E}}(t,\phi), v_r)  =  \mu(I_{\lambda \phi} I_\text{tot,H}, v_r) 
\label{eq:solarProfileVsDistanceH}
\end{eqnarray}
and the normalization factor $p_\text{H}$ is 
\begin{eqnarray}
p_\text{H} & = & \left[f_{\text{osc},\text{H}} \left(\frac{\pi\, e^2}{m_e c^2}\, h\, \lambda_\text{H} \right)\frac{r\E^2}{G\,M_\sun m_\text{H}}\right]^{-1} 
 =  3.34467\times 10^{12} \text{ ph s}^{-1}\,\text{cm}^{-2}\,\text{nm}^{-1}.
\label{eq:normFactHDefinition}
\end{eqnarray}
For derivation, see Appendix \ref{sec:radPress}. To obtain the illuminating solar spectral flux, the quantities $s_\text{H}$ and $p_\text{H}$ thus defined must be inserted into Equation~\ref{eq:solSpectrFluxDef}. 

The left panel of Figure~\ref{fig:lineProfiles} presents profiles of $s_\text{H}$ for the solar minimum and maximum conditions. The two profiles are linearly scaled so that they are equal to 1 for $v_r = 0$. This illustrates the variation of the shape of the solar line profile with $I\tot$, which is more complex than a simple linear scaling. In particular, the horn/center ratios vary during the cycle of solar activity, thus the illuminating solar spectral flux $\tilde{S}_\sun$ varies during the solar cycle both in the absolute magnitude and in the relation of the flux at one wavelength to another. 

\subsubsection{The spectral profile for helium $s_\text{He}$ and the scaling factor $p_\text{He}$}
\label{sec:solarHeLineProfile}
\noindent
The central wavelength of the He I 58.4 nm line is equal to $\lambda_{\text{He}} = 58.43339$ nm, and the oscillator strength $f_{\text{osc},\text{He}} = 0.27625$ \citep{wiese_fuhr:09a}. 
The solar He~I~58.4 nm line is so narrow that available measurements of its profile are few and far between (see Appendix~\ref{sec:appendix1}). It is expected that the profile is Gaussian-like with a flat spectral background or kappa-like \citep[e.g.,][]{jeffrey_etal:17a}. In our model, the Full Width at Half maximum (FWHM) of the profile was adopted as $\text{FWHM}_\lambda = 0.0118$~nm, independent of the phase of the solar cycle \citep{mcmullin_etal:04a}, which corresponds to FWHM$_v = 60.54$ \kms. A rationale behind this choice and a discussion on available measurements of this line is provided in Appendix \ref{sec:appendix1}.

We define a normalized Gaussian line profile with a background as follows:
\begin{equation}
s_{\text{He}}(v_r, \vD, s_\text{bkg}) = \frac{1}{1 + s_{\text{bkg}}}\left(\exp \left[-\left(\frac{v_r}{\vD} \right)^2\right]  +s_{\text{bkg}}\right).
\label{eq:defHe584LineProfile}
\end{equation}
The profile is parametrized by a Doppler speed, a Doppler width, and a background level. The relation between the line width expressed by the central wavelength $\lambda$, line width  $\Delta \lambda$, and Doppler speed $v_\text{D}$ is 
\begin{equation}
\vD = (\Delta \lambda/\lambda) c,
\label{eq:defineDopplerShift}
\end{equation}
where $c$ is the speed of light. 

It is expected that the $s_{\text{bkg}}$ is small relatively to the spectral irradiance at the line center. In the definition in Equation~\ref{eq:defHe584LineProfile}, the spectral irradiance at the central wavelength is normalized to 1. The variation in the intensity of the solar line is accounted for so that the profile maintains its shape and is scaled linearly with $I_{\text{tot,He}}(t)$.

For $s_{\text{bkg}} = 0$, the relation between the full width at half maximum (FWHM$_v$) and $\vD$ is given by the formula:
\begin{equation}
\vD = \frac{\text{FWHM}_v}{2 \sqrt{\ln 2} }. 
\label{eq:gaussFWHMVsVd}
\end{equation}
For $0 < s_{\text{bkg}} < 1$, this relation becomes
\begin{equation}
\vD = \frac{\text{FWHM}_v}{2 \sqrt{\ln \frac{2}{1-s_\text{bkg}}} }.
\label{eq:gaussFWHMVsVdBkg}
\end{equation}

Alternatively, one can define the profile by a kappa function:
\begin{equation}
s_{\text{He}}(v_r, \vD, \kappa) = \left[1+\frac{1}{\kappa}\left(\frac{v_r}{\vD}\right)^2 \right]^{1-\kappa},
\label{eq:defKappaProfile}
\end{equation}
where \vD{} of the profile is related to FWHM$_v$ by
\begin{equation}
\vD = \frac{\text{FWHM}_v}{2 \sqrt{\kappa \left(2^{\frac{1}{\kappa-1}} - 1 \right)}}.
\label{eq:defKappaFWHM}
\end{equation}
In fact, as illustrated in the right panel of Figure~\ref{fig:lineProfiles}, a Gaussian profile with a moderate background differs from a kappa profile with identical FWHM only in the tails even for relatively low magnitudes of the $\kappa$ parameter. 

The definitions presented above include a normalization with $s_\text{He}(v_r = 0) = 1$. In the calculations, we have the magnitude of the line-integrated flux $I_{\text{tot}}$ corresponding to a certain waveband $\lambda_1, \lambda_2$; $\lambda_2 - \lambda_1 = 1$ nm, i.e., within a nanometer-wide wavelength interval that includes the central wavelength of the He I 58.4 nm line. The boundaries $(\lambda_1, \lambda_2)$ can be expressed by the Doppler velocities $(v_{r,1}, v_{r,2})$ (Equation \ref{eq:freqVrDefinition}). The normalization factor $p_\text{He}$ for the line profile must satisfy the following condition:
\begin{equation}
I_{\sun,\text{E}} = I_{\lambda \phi}(\phi) I\tot(t) = p_\text{He} \int \limits_{v_{r,1}}^{v_{r,2}} s_{\text{He}}(v_r)\, d v_r, 
\label{eq:normFactHeDefinition}
\end{equation}
where $s_{\text{He}}(v_r)$ is the function defined in Equation \ref{eq:defHe584LineProfile} or \ref{eq:defKappaProfile} with the relevant parameters invariable with time.

Ultimately, the illuminating solar spectral function 
\begin{equation}
\tilde{S}_{\sun,\text{He}}(v_r,t) = p_\text{He}( I_{\text{tot,He}})\, s_\text{He}(v_r),
\label{eq:sSunHeDef}
\end{equation}
where, unlike for H, the factor $p_\text{He}$ (Equation \ref{eq:normFactHeDefinition}) depends on the line-integrated solar flux $I_\text{tot,He}$, and consequently on time; $I_{\lambda \phi}$ is given by Equation \ref{eq:iSunPhiDefinition}.

\subsubsection{Modulation of the illuminating solar flux with heliolatitude $I_{\lambda \phi}$}
\label{sec:iLambdaPhi}
\noindent
The solar EUV output is a superposition of components from the quiet Sun, the low corona, and  all kinds of active regions \citep{amblard_etal:08a}. These latter ones are mostly responsible for the variation during the solar activity cycle and solar rotations. The fact that this component originates in discrete regions on the solar disk, which tend to cluster in heliolatitudinal bands, results in a latitudinal and longitudinal variation of the line-integrated solar flux in space $I_\sun$, evolving with time. 

The dimensionless factor $I_{\lambda \phi}$ is a modulating function, representing variations of the line-integrated solar flux with the heliolatitude and the heliolongitude. This latter effect is caused by solar active regions, located at the surface of the rotating Sun. 

The variation of $I_{\lambda \phi}$ in the longitude was pointed out in the context of the helioglow by \citet{bertaux_etal:00, pryor_etal:92} as responsible for the ``searchlight effect'', i.e., an apparent motion of a reflection of an active region in the sky. The Sun features a rotation-related flickering in the \lya line with the statistically most likely magnitude of $\sim 7$\%, as we derived from the daily time series of $I_\text{tot}$ by \citet{machol_etal:19a}, restricted to a portion of these data that originate directly from \lya{} measurements, and not from proxies. The median value of the maximum to minimum ratio within individual Carrington rotations is equal to 9\% and the mean to 11\%, while magnitudes larger than 24\% are detected during 3\% of days in the discussed time series. These variations are reflected in brighter sky regions several dozens of degrees across. The magnitude of variations in the helioglow intensity observed along selected lines of sight depends on the viewing geometry and is the largest for antisolar lines of sight. For lines of sight close to perpendicular to the Sun-observer line, they are typically 5--10\% percent, as obtained from analysis of selected SWAN data by \citet{bzowski_etal:03a}. Accounting for the searchlight effect is easiest for the viewing geometries close to anti-solar; this is because for these geometries, the largest portion of the line of sight is affected by the active region-related brightening. In this case it seems sufficient to use for illumination a daily value of $I_\text{tot}$ instead of the monthly-averaged one, which is used to calculate the distribution of ISN H.

In the present version of our model, we neglect this variation and assume that the modulation factor is only a function of the heliolatitude and does not vary with time: $I_{\lambda \phi}(t, \lambda_\text{helio}, \phi_\text{helio}) \equiv I_{\lambda \phi}(\phi_\text{helio})$. In the following, we will abbreviate $\phi_\text{helio}\equiv \phi$. We define this function requiring a normalization such that 
\begin{equation}
I_{\lambda \phi}(\phi = 0) = 1,
\label{eq:iLambdaPhiNormalization}
\end{equation} 
i.e., it is equal to 1 at 1 au at 0 heliolatitude, i.e., close to the latitude where measurements of $I_\sun$ are taken. 

The dependence of the solar EUV output on heliolatitude has not been fully investigated \citep[see discussion in][]{bzowski_etal:12b}. Studies by \citet{pryor_etal:92} and \citet{auchere:05} suggest that the intensity of solar EUV at polar latitudes is somewhat lower than in the ecliptic, and that this difference changes little with the solar activity. \citet{pryor_etal:92} considered a superposition of contributions from the quiet Sun, assumed to be spherically symmetric, and another one from active regions at the solar surface, organized by heliolatitude. Effectively, the heliolatitude dependence of $I_{\sun,\text{E}}(\phi)$ in this approach was given by a constant spherically symmetric portion and a portion with a certain amplitude, modulated as $\cos^2 \phi$, i.e., $I_{\sun,\text{E}}(\phi) = c_1 + c_2 \cos^2 \phi$. These authors connected the $c_1$ and $c_2$ parameters with the intensity for the quiet Sun and a portion related to the latitude distribution of active regions and their surface density. These parameters can be obtained from solar disk observations. This approach is worth a more thorough analysis but in our paper we use a formula with a reasonable but arbitrary parameter value because we only want to investigate the sensitivity of the helioglow to an anisotropy in the illumination. To that end, we tentatively adopt the following formula:
\begin{equation}
I_{\lambda \phi}(\phi) = a\, \sin^2\phi  +  \cos^2 \phi,
\label{eq:iSunPhiDefinition}
\end{equation}
where $a = I_{\text{pole}}/I_{\text{tot}}$ (note that from Equation \ref{eq:iTotVsISun}, $I_{\text{tot}}=I_{\text{eqtr}}$ ) is an adjustable parameter, equal to 0.85, both for H and He. This function, however, can and most likely will be redefined in the future, when a better insight is available. Setting $a = 1$ removes the latitude dependence of the solar illumination in the model. Note that for the \citet{pryor_etal:92} formula, $I_\text{pole}/I_\text{tot} = c_1/(c_1 + c_2)$. It can be shown using simple algebra that substituting $a = c_1/(c_1 +c_2)$ to $I_{\sun,\text{E}}=I_\text{tot}\,I_{\lambda \phi} = I_\text{tot} (a \sin^2 \phi + \cos^2 \phi)$ one obtains $I_{\sun,\text{E}}(\phi) = (I_\text{tot}/(c_1 + c_2))(c_1 + c_2\cos^2 \phi)$.

\subsubsection{The total illuminating solar flux $I_{\sun}$}
\label{sec:totalIntensity}
\noindent

The total illuminating solar flux  $I_\sun(t, \myvec{r}\los)$ measured at $\myvec{r}\los$ is integrated over the solar surface hemisphere defined by the unit vector $\hat{\myvec{r}}\los$. The flux $I_\sun$ varies quadratically with the distance from the Sun (assuming no absorption). In the paper, we assume that for a time $t$, the total solar flux is known at a reference distance $r\E = 1$~au at the solar equator ($\phi = 0$) and we denote this quantity 
\begin{equation}
I\tot(t) = I_\sun(t, r=r\E, \phi = 0).
\label{eq:iTotVsISun}
\end{equation}
This quantity is measured by Earth-orbiting satellites for H and He \citep{woods_etal:05a, woods_etal:12a}. The total illuminating flux $I_\sun(t, \myvec{r}\los)$ at $\myvec{r}\los$ in the spherical heliographic coordinates is then given by 
\begin{equation}
I_{\sun}(t,\myvec{r}\los) = I_{\sun,\text{E}}(t,\phi) \left (\frac{r\E}{r\los}\right)^2 = I_{\lambda \phi}(\phi) I\tot(t) \left(\frac{r\E}{r\los}\right)^2,
\label{eq:IsunDefinition}
\end{equation}
that is, the variation of the solar output $I_\sun(t, \myvec{r}\los) $ far away from the Sun $(r\los \gg r_{\sun})$ is equal to the solar output $I_{\sun,\text{E}}(t, \phi)$ measured at a heliolatitude $\phi$ and a reference distance $r\E$, and scaled with the square of solar distance. $I_{\sun, \text{E}}$ is factorized into a (heliolongitude, heliolatitude) variation $I_{\lambda, \phi}$ and the global time dependent part $I\tot(t)$ measured at the solar equator. 

In reality, the solar flux is attenuated by the intervening gas. This phenomenon was recently discussed by \citet{IKL:18b} -- see their Figures 8 and 9. Since absorption is wavelength dependent, both the magnitude and the spectral profile of the solar illuminating flux at $\myvec{r}\los$ must vary with $\myvec{r}\los$ (differently for different heliolongitudes, heliolatitudes and solar distances). 
Therefore, while the solar \lya and 58.4 nm line profiles at $r\E = 1$~au can be assumed to be identical to those just outside the solar corona (because of the negligible optical thickness of the ISN gas inside 1 au), outside 1 au they are gradually modified. 
Consequently, the illuminating solar spectral flux profile is a function of wavelength and the location along the line of sight. The effect of absorption will be introduced to a future version of the model, but to the function $s$ describing the solar line profile rather than to the function of the total illuminating flux.

The time variations in the ecliptic plane have been demonstrated to be highly correlated with the well-known proxies of the solar activity. The daily \lya flux measured at the Earth location is publicly available in the form of so-called solar composite \lya flux \citep{woods_etal:96a, woods_rottman:97, woods_etal:05a}, with the most recent version given by \citet{machol_etal:19a}. This data product is based on observations from space from several spacecraft since 1970s, with gaps filled with the proxies \citep[see][]{woods_etal:00, bzowski_etal:12b}. This product is adopted as the basis for the solar illuminating flux $I_\text{tot,H}(t)$ for hydrogen. We use the original time series, available at a daily time resolution, averaged over Carrington rotation period to average over variations in heliolongitude and we assume that for a given time moment $t$ the Carrington period-averaged $I\tot(t)$ is identical for all heliolongitudes. 

The solar 58.4 nm line for helium, $I_\text{tot,He}(t)$, has been measured by SDO \citep{woods_etal:12a} and also features correlations with the solar proxies. We adopted $I_\text{tot,He}$
as the data product from SDO/EVE level 3 lines dataset version 6 provided by LASP Interactive Solar Irradiance Datacenter (\url{http://lasp.colorado.edu/lisird/data/sdo_eve_lines_l3/}), averaged over Carrington period identically as in the case of hydrogen. 

\subsection{Numerical integration of the source function to yield the helioglow intensity $I\glow$}
\label{sec:numIntegr}
\noindent
Combination of Equations 
\ref{eq:helioglowIntensDef1}, \ref{eq:helioglowSourceFunDef}, \ref{eq:defSrcFun}, \ref{eq:partialDens}, \ref{eq:solSpectrFluxDistance}, \ref{eq:excitFunWorkDef2}, \ref{eq:solSpectrFluxDef} and  \ref{eq:IsunDefinition} 
results in the following integration formula:
\begin{eqnarray}
I\glow(\myvec{r}\obs, \myvec{\hat{l}}) & = & \int\limits_0^{\infty} \int \limits_{-\infty}^{\infty} E(\tilde{S}_\sun, \myvec{l}, v_r)\, \eta(\myvec{l}, v_r)\, \psi(\beta)\, d v_r \,dl = \nonumber \\ 
                                  & = &
\frac{\pi\,e^2}{m_e\, c}\, f_{\text{osc}} \, p
\int \limits_0^{\infty} d l \,
\psi(\beta)\, \left(\frac{r\E}{r\los}\right)^2  
\label{eq:helioGlowIntensityDef} \\
& & \int \limits_{-\infty}^{\infty} d v_{T_1} 
\int \limits_{-\infty}^{\infty} d v_{T_2}  
\int \limits_{-\infty}^{\infty} d v_r \, s(I_{\sun,\text{E}}(t,\phi),v_r)
f_\text{ISN}(\myvec{l}, v_r, v_{T_1}, v_{T_2}). \nonumber 
\end{eqnarray}
Note that the result of integration in the last row of this equation for hydrogen is proportional to the radiation pressure averaged over the entire population at the location, given by the radius vector $\myvec{l}$ within the line of sight. The result of this integration is a function of this location, and the result, of course, does not depend on the details such as selection of the coordinate system. Therefore, it is not needed to change the coordinate system along the line of sight such to always keep one of the axes heliocentric. In our WawHelioGlow code, the integration is performed in the spherical heliographic coordinates, to facilitate the use of latitudinally anisotropic models of ISN gas ionization and of the illuminating solar flux $I_{\lambda \phi}$. We perform a four-dimensional numerical integration of the helioglow intensity $I$ according to Equation \ref{eq:helioGlowIntensityDef}, and to calculate the integrand function $f_\text{ISN}(\myvec{l}, v_r, v_{T_1}, v_{T_2})$, we perform tracking of an individual atom trajectory by calculation of numerical solution of the equation of motion of a H atom between $\myvec{l}$ until this atom reaches a predefined distance from the Sun (typically selected about 150 au). 

If we wanted to calculate an average radiation pressure for the sample of H atoms at $\myvec{l}$, we would need to calculate 
\begin{equation}
\langle\mu(\myvec{l}) \rangle = \frac{\int \mu(I_{\sun,\text{E}},v_r) f_\text{ISN}(\myvec{l},\myvec{v})\, d^3 v}{\int f_\text{ISN}(\myvec{l},\myvec{v})\, d^3 v}.
\label{eq:averageMu}
\end{equation} 
The term in the denominator is equal to the local density $n$, so effectively for H, Equation \ref{eq:helioGlowIntensityDef} can be written down as:
\begin{eqnarray}
I\glow(\myvec{r}\obs, \myvec{\hat{l}}) & = &
\frac{\pi\,e^2}{m_e\, c}\, f_{\text{osc,H}} \, p_\text{H}
\int \limits_0^{\infty} d l \,
\psi(\beta)\, \left(\frac{r\E}{r\los}\right)^2\, n (\myvec{l}) \, \langle \mu(\myvec{l})  \rangle.
\label{eq:helioGlowIntensityH}
\end{eqnarray}

In the numerical calculations, we integrate over $l$ up to a limiting distance $L\approx 60$ au for H, and $L\approx 20$ au for He, and add a correction term, which accounts for the integration to $\infty$:
\begin{equation}
\begin{aligned}
I\glow(\tilde{S}_\sun, \myvec{r}\obs, \hat{\myvec{l}}) = 
\int\limits_{0}^L J\glow(\tilde{S}_\sun, \myvec{l})\, d l  +  I_\text{rest}, 
\\
I_\text{rest}=\int\limits_{L}^{\infty} J\glow(\tilde{S}_\sun, \myvec{l})\, d l .
\end{aligned}
\label{eq:corrInfLineOfSign}
\end{equation}
 
Because for large distances from the Sun we assume $l\approx r\los$ and using Equation \eqref{eq:solSpectrFluxDistance} we obtain $ J\glow(\tilde{S}_\sun,l)= J\glow(\tilde{S}_\sun,L)(r\E/l)^2$. After analytic integration of $\int 1/l^2\, dl$ we have 
\begin{equation}
I_\text{rest}= -\frac{J\glow(\tilde{S}_\sun,L)\,r\E}{l}\Big|_L^{\infty} = \frac{J\glow(\tilde{S}_\sun, \myvec{L})\,r\E}{L}.
\label{eq:intensityRest}
\end{equation}
Selection of $L \approx 60$~au and $L \approx 20$~au is based on our experience with test runs of the simulation program for observer locations at about 1 au from the Sun. The integration defined in Equation \ref{eq:helioGlowIntensityDef} is performed using an adaptive-step numerical method that increases the numerical integration step when it determines that the increment of the integrand function becomes small. Consequently, at 20 au and 60 au from the Sun the integration step may be quite large. To save the calculation time, we use a loose criterion for finishing the numerical part of the integration, requiring that the routine exceed the $L$ limit but need not achieve it precisely. At these distances from the Sun the lines of sight originating at 1 au run very close to the radial direction, and radial gradients of the density, velocity, and thermal spread of ISN gas are very small, which makes the approximation defined in Equation \ref{eq:corrInfLineOfSign} very reasonable. Nevertheless, the correction term $I_\text{rest}$ in most cases cannot be neglected in comparison with our target numerical accuracy of $\approx 2$\%.

\section{Summary and outlook}
\label{sec:summaryOutlook}
\noindent
In this paper, we present Warsaw Heliospheric Glow model (WawHelioGlow) -- a numerical code for simulating the helioglow intensity of ISN H and He. The model, based on the (n)WTPM model of the ISN gas distribution in the heliosphere, takes into account the time evolution and heliolatitudinal dependence of the solar factors: spectral illumination, radiation pressure, and ionization losses of the ISN gas inside the heliosphere. The evolution of these factors in the simulations is included by using well-established models based on relevant observations. The construction of the code facilitates replacing these models with alternative ones.

Presented is the first version of the WawHelioGlow code, which uses an optically thin, single scattering approximation of radiation transfer. However, the distribution function of the ISN gas is simulated fully numerically, without the frequently used approximation of Maxwell-Boltzmann distribution function for the ISN gas inside the heliosphere. Including effects of multiple scattering of the solar \lya photons will be implemented in a future version of the model. 

Effects of multiple scattering were discussed by several authors, e.g., \cite{ajello_etal:94, hall:92a, keller_thomas:79a, quemerais_bertaux:93a, quemerais:00, quemerais_etal:19a, scherer_fahr:96}. Their magnitude depends on details of approximations made in the multiple scattering approach, in particular on the model of frequency redistribution. Multiple-scattering codes are notorious for their high demand on computer power. \citet{ajello_etal:94} and \citet{quemerais:00} suggested pre-calculating the ratio of multiple-scattering and optically thin approximations and using them as correction terms in further analyzes of observations using single-scattering modeling. The corrections are the largest for the downwind direction, on the order of 1.25 \citep[see Table 5 in ][]{quemerais:00}. This approach seems to work best for viewing geometries close to anti-solar.

The multiple scattering correction to an optically thin model is not an optimum solution especially for LOS geometries much different from antisolar because the additional excitation of the gas due to the scattered light is basically a function of the angular distance from the upwind direction and the distance from the Sun; when the line of sight traverses a wide range of offset angles, then it is not straightforward to decide on the magnitude of correction to apply even for observer locations at 1 au. Clearly, it is best to have a multiple-scattering code. WawHelioGlow will account for multiple scattering effects in one of its future versions.

Another effect that will be addressed in a future version of the code is the 27-day variation associated with active regions at the Sun's surface \citep{pryor_etal:92, pryor_etal:96, bertaux_etal:00}. Accounting for this effect is easiest in the viewing geometries close to anti-solar. The amplitude of the variation is largest for these geometries; this is because for these geometries, the largest portion of the line of sight is affected by the active region-related brightening. In future versions of the code it is planned to use daily values of the solar flux $I_\text{tot}$ for illumination of the gas and Carrington rotation-averaged $I_\text{tot}$ values for the calculation of the ISN H distribution. 

The WawHelioGlow model includes provisions to account for a heliolatitudinal anisotropy in the solar EUV output. This anisotropy is poorly investigated but, based on available evidence, must be regarded as a distinct possibility. The EUV anisotropy affects on one hand the illumination of the gas, and on the other hand the spatial distribution of the ISN H and He populations because of their sensitivity to the radiation pressure (only H) and photoionization. For H, this anisotropy is a secondary anizotropization effect, in addition to the primary effect of the solar wind anisotropy impressed in the gas by charge exchange. For He, the EUV anisotropy is the main anizotropization factor because He is the most sensitive to photoionization. In the code, the anisotropy is currently implemented in a simplified way as a single-parameter analytical formula, but it can be easily replaced with a more refined model.

The spectral illumination in the code can be defined in a variety of ways. The code does not rely on any particular analytic definition or normalization of the profile, the solar line profiles can be defined arbitrarily. In the present version, for He it implements a Gaussian profile with a uniform spectral background or, alternatively, a kappa profile. In the case of hydrogen, the present version uses a state of the art model of the evolution of the solar \lya line. This model is based on spectral measurements performed in the ecliptic plane throughout the cycle of the solar activity \citep{lemaire_etal:15a}. It assumes that the spectral shape depends solely on the total \lya flux. In connection with the model of the anisotropy, this results in an extrapolation of the measurement-based spectral model towards low values of the total solar flux during low solar activity at high heliolatitudes. This feature of the WawHelioGlow model must be regarded as an educated speculation. However, it is easy to implement a different approach and test the results against observations. 

The WawHelioGlow model is optimized towards analysis of future observations of the helioglow by the GLOWS experiment onboard the planned NASA IMAP mission. It is also well suited for analysis of photometric observations of the helioglow from the SWAN experiment onboard SOHO, as well as other photometric observations of the helioglow of both H and He, performed within a few au from the Sun. 

A presentation of the behavior of the factors making up the helioglow signal for various observation geometries and phases of the solar cycle, for H and He, is given in the accompanying paper. 
That paper also shows a qualitative comparison of selected maps of the \lya helioglow observed by SWAN during a minimum and a maximum of the solar activity. To these comparisons we used the results of the WawHelioGlow code obtained with the use of a state of the art model of the ionization factors and solar illumination, and either including or excluding the model of the latitudinal anisotropy of the solar EUV output. That paper illustrates that WawHelioGlow model will potentially be able to represent the observed helioglow intensity using the existing models of the evolution of the solar ionization factors when latitudinal variation of the solar EUV output is better investigated. 

\begin{acknowledgments}
{\emph{Acknowledgments}}. This study was supported by Polish National Science Center grants 2019/35/B/ST9/01241, 2018/31/D/ST9/02852, and by Polish Ministry for Education and Science under contract MEiN/2021/2/DIR.
\end{acknowledgments}

\appendix
\section{Cross section for photon absorption, the G factor, and radiation pressure}
\label{sec:appendixCrossSection}
\subsection{Introduction}
\label{sec:units}
\noindent
In this appendix, we present textbook derivations of some of the quantities relevant for modeling the helioglow, the resonant absorption and radiation pressure, and related quantities. This is a {\emph{pro memoria}} description, also intended as a quick-reference source. 

Throughout the helioglow work, we use a common base of the numerical values of physical quantities. To facilitate maintaining this system, in Table \ref{tab:units} we present a list of these physical constants and their units in the cgs system. We will use them to derive normalization factor $p_\text{H}$ that allows us to convert between radiation pressure profile in the so-called $\mu$-units and the solar spectral flux in the physical units, as well as the $g$-factor, informing on the frequency of atom excitation at a given distance to the Sun. 
\begin{deluxetable}{llll}
\tablecaption{\label{tab:units} Physical constants in cgs units. In our numerical simulations, we use more precise values based on the NIST standards.
}
\tablehead{\colhead{constant} & \colhead{symbol} & \colhead{value} & \colhead{unit}}
\startdata
elementary charge  & $e$   & $4.8032 \times 10^{-10}$ & $ \text{cm}^{3/2} \text{ g}^{1/2} \text{ s}^{-1}$\\
electron mass     & $m_\text{e}$ & $9.1094 \times 10^{-28}$ & g \\
speed of light    & $c$     & $2.9979 \times 10^{10}$  & $ \text{cm s}^{-1}$\\
\lya wavelength & $\lambda_\text{H}$ & $ 121.567$ & nm\\
He I wavelength & $\lambda_\text{He}$ & $58.433 $ & nm\\
gravitational constant & $G$ & $6.6743 \times 10^{-8}$ & $ \text{cm}^{3} \text{ g}^{-1} \text{ s}^{-2}$\\
astronomical unit & $r_\text{E}$ & $1.496 \times 10^{13}$ & cm\\
solar mass & $M_\sun$ & $1.989 \times 10^{33}$ & g\\
hydrogen mass & $m_\text{H}$ & $1.6726 \times 10^{-24}$ & g\\
Gaus gravity constant & $k^2$ & $2.959122130672713\times 10^{-4}$ & au$^3$ day$^{-1}\, M_\sun$\\
spectral flux & $\pi F_\lambda$ & time dependent & $\text{ph cm}^{-2} \text{s}^{-1} \text{ nm}^{-1}$\\
radiant intensity & $I_\nu$ & time dependent & $\text{erg s}^{-1} \text{cm}^{-2} \text{ sr}^{-1} \text{ Hz}^{-1}$\\
H oscillator strength & $f_{\text{osc},\text{H}}$ & $0.41641$ & dimensionless\\
He oscillator strength & $f_{\text{osc},\text{He}}$ & $0.27625 $ & dimensionless\\
combination & $\frac{\pi \text{e}^2}{m_\text{e} \text{c}}$ & $0.0265402$ &$\text{cm}^{2} \text{ s}^{-1}$\\        
\enddata
\end{deluxetable}

\newpage
\subsection{Cross section}
\label{sec:crossSectn}
\noindent
Neutral atoms in the heliosphere are exposed to EUV radiation from the Sun. Sometimes, they absorb a photon and then immediately re-emit it in a different direction and with a slightly different frequency. The process must conserve the total momentum of the atom and the photon. As a result, due to the photon absorption the momentum of the atom is incremented by that of the photon. In one act of absorption, the momentum of the atom changes by:
\begin{equation}
\Delta \text{p}=\frac{\text{h}}{\lambda}
\label{eq:deltaP}
\end{equation}
and the direction of the momentum increment is anti-solar. After re-emission, the momentum changes by a very similar amount in a random direction, with the likelihood proportional to the phase function (Equations \ref{eq:defPhaseFunH}, \ref{eq:defPhaseFunHe}). Since, however, the momentum changes due to re-emission average to zero change because the scattering phase function is axially symmetric about the photon impact direction, the momentum change due to re-emission is neglected in the further discussion.

A measure of how often photons are absorbed by the gas atoms is the cross section for absorption. The absorption is considered in the inertial reference frame of the atom. In the following, we derive the cross section using the classical theory. Fundamentally, the same result can be obtained using the quantum theory, which returns the correct species and transition-dependent result. The classical result must be scaled to the quantum-mechanical one using a proportionality constant referred to as the oscillator strength $ \text{f}_{\text{osc}}$ for the given atom and transition.

Considering an atom in a given quantum state as a classical damped oscillator with radiation as an exciting force, we find that the cross section can be expressed as:
\begin{align}
\label{eq:crossSec1}
\sigma(\omega)&=\frac{8\pi\text{e}^4\omega^4}{3\text{m}_\text{e}^2 c^4}\frac{1}{(\omega^2-\omega_0^2)^2\gamma^2\omega^2},\\
\gamma&=\frac{2\text{e}^2 \omega_0^2}{3\text{m}_\text{e} \text{c}^3},
\end{align}
where $\omega$ is the eigenfrequency, and $\gamma$ is the damping constant of the considered oscillator.

Since we consider frequencies near the resonance ($\Delta \omega=\omega-\omega_0 \ll \omega_0$), we can apply the following approximation:
\begin{equation}
\omega^2-\omega_0^2=(\omega-\omega_0)(\omega+\omega_0) \approx 2\omega_0(\omega-\omega_0).
\end{equation}  
Every other appearance of $\omega$ in Equation \ref{eq:crossSec1} can be substituted by $\omega_0$:
\begin{equation}
\label{eq:crossSec2}
\sigma(\omega)=\frac{2\pi\text{e}^2}{\text{m}_\text{e} \text{c}}\frac{\frac{\gamma}{2}}{(\omega-\omega_0)^2(\frac{\gamma}{2})^2}.
\end{equation}
This equation is very convenient to interpret. The second part is known as the Lorentzian function that has very well defined properties, the first part is a proportionality constant.

Equation \ref{eq:crossSec2} can be expressed either as a function of the radiation frequency or as a function of its wavelength:
\begin{align}
\Delta \omega&=2\pi \Delta \nu\\
\Delta \omega&=\frac{2 \pi \text{c}}{\lambda^2} \Delta \lambda.
\end{align}

With this, we can write the following formulae expressing the spectral cross section (dependent on the frequency or alternatively the wavelength) and the total cross sections, integrated over the frequency.
In the reality, atoms should be described using quantum mechanics rather than the classical theory.
Therefore, in Equations \ref{eq:crossSecF1}---\ref{eq:intCrossSecW1} we introduced the oscillator strength $\text{f}_{\text{osc}}$, which is a scaling factor that makes the classical result in agreement with the quantum-mechanical one.
\begin{align}
\label{eq:crossSecF1}
\sigma(\nu)&=\frac{\text{e}^2}{\text{m}_e \text{c}}\text{f}_{\text{osc}}\frac{\frac{\gamma}{4 \pi}}{(\nu-\nu_0)^2(\frac{\gamma}{4 \pi})^2},\\  \nonumber
\label{eq:intCrossSecF1}
\sigma_{\nu}=\int_0^{\infty}d\nu \sigma(\nu)&=\int_{-\infty}^{\infty}d(\Delta \nu)\sigma(\Delta \nu)\\ \nonumber
&=\frac{\text{e}^2}{\text{m}_e \text{c}} \text{f}_{\text{osc}} \int_{-\infty}^{\infty}d(\Delta \nu) \frac{\frac{\gamma}{4 \pi}}{(\Delta \nu)^2(\frac{\gamma}{4 \pi})^2}\\
&=\frac{\pi \text{e}^2 }{\text{m}_e \text{c}} \text{f}_{\text{osc}} & [\text{cm}^2\text{Hz}],
\end{align}
 or the wavelength:
\begin{align}
\label{eq:crossSecW1}
\sigma(\lambda)&=\frac{\text{e}^2}{\text{m}_e \text{c}}\frac{\lambda_0^2}{\text{c}}\text{f}_{\text{osc}}\frac{\frac{\gamma \lambda_\text{0}^2}{4 \pi \text{c}}}{(\lambda-\lambda_\text{0})^2(\frac{\gamma\lambda_\text{0}^2 }{4 \pi \text{c}})^2}\\ \nonumber
\label{eq:intCrossSecW1}
\sigma_{\lambda}=\int_0^{\infty}d\lambda \sigma(\lambda)&=\int_{-\infty}^{\infty}d(\Delta \lambda)\sigma(\Delta \lambda)\\ \nonumber
&=\frac{\text{e}^2}{\text{m}_e \text{c}}\frac{\lambda_0^2}{c} \text{f}_{\text{osc}} \int_{-\infty}^{\infty}d(\Delta \lambda) \frac{\frac{\gamma \lambda_\text{0}^2}{4 \pi \text{c}}}{(\Delta \lambda)^2(\frac{\gamma \lambda_\text{0}^2}{4 \pi \text{c}})^2}\\
&=\frac{ \pi \text{e}^2}{\text{m}_e \text{c}} \frac{\lambda_0^2}{c} \text{f}_{\text{osc}}  & [\text{cm}^2\text{nm}].
\end{align}
The oscillator strength can be calculated theoretically as well as directly measured.

\subsection{Factor g}
\label{sec:gFactor}
\noindent
The so-called g factor is a measure of how many photons are absorbed by an atom in a unit time given an incident spectral flux. For an atom stationary relative to the Sun, it can be computed in following way:
\begin{equation}
\label{eq:g1}
\text{g}=\int_0^{\infty}d\lambda \pi \text{F}(\lambda_0) \sigma(\lambda),
\end{equation}
where $\pi \text{F}(\lambda_0)$ is the solar spectral flux in the center of the line (defined by wavelength $\lambda_0$) expressed in the units $\text{ph cm}^2 \text{s}^{-1}\text{nm}^{-1}$. 

For the \lya radiation, which is relevant for H atoms, we do have direct measurements of the spectral flux in the center of the line \citep{lemaire_etal:05}. In the case of the He I line, measurements of the profile are very few and far between (see Appendix \ref{sec:appendix1}) and on a longer basis, there are only measurements of the total flux integrated over whole line available \citep{woods_etal:15a}. This is because the He I line is much narrower than the \lya line. Therefore, we will follow the approach presented in Equation 4 of \citet{grava_etal:18a} and calculate the flux in the center of line assuming the Gaussian shape. Either way, the flux $\pi \text{F}(\lambda_0)$ is a wavelength-independent value that can be taken out of the integral.
Then, using Equation \ref{eq:intCrossSecW1} and Equation \ref{eq:g1}, we can write a simple formula for the g factor:
\begin{align}
\label{eq:gfactor}
\text{g}&=\pi \text{F}(\lambda_0) \frac{\pi \text{e}^2}{\text{m}_e \text{c}} \frac{\lambda_0^2}{\text{c}} \text{f}_{\text{osc}} & [\text{ph atom}^{-1} \text{s}^{-1}].\\ \nonumber
\lambda_0&=
\begin{cases}
   \lambda_\text{H}  & \text{Hydrogen} \\
   \lambda_\text{He} & \text{Helium}.
\end{cases}\\ \nonumber
\text{f}_{\text{osc}}&=
\begin{cases}
   \text{f}_{\text{osc,H}}  & \text{Hydrogen} \\
   \text{f}_{\text{osc,He}} & \text{Helium}.
\end{cases}\\ \nonumber
\pi \text{F}(\lambda_0)&=
\begin{cases}
  \text{I}_\text{Lya}( \lambda_\text{H}) & \text{Hydrogen} \\
  \frac{ \text{I}_\text{tot,He} \frac{\lambda_\text{He}}{\text{hc}} 10^3 }{1.064 \text{ FWHM}_{\lambda}}  & \text{Helium}
    \end{cases}\\
\end{align}
where $\text{I}_\text{Lya}( \lambda_\text{H})$ is the spectral flux in the center of the \lya line expressed in units of $\text{ph cm}^2 \text{s}^{-1}\text{nm}^{-1}$, $\text{I}_\text{tot,He}$ is the total flux in He I line expressed in units of W m$^{-2}$, $\text{FWHM}_\lambda=0.0136$ nm (see Appendix \ref{sec:appendix1}) is the Full Width at Half Maximum of the He~I line assuming its Gaussian shape, the $10^3$ factor is for conversion between units. For the product $G\,M_\sun$ it is best to use the Gauss gravity constant squared: $G\,M_\sun = k^2$ (see Table~\ref{tab:units}). This is because this quantity is much more precisely known than the gravitational constant and the solar mass separately.

For an atom with a non-zero radial velocity $v_r$ relative to the Sun, the spectral flux at the line center in Equation \ref{eq:gfactor} must be replaced with the spectral flux corresponding to the Doppler-shifted wavelength given by Equation \ref{eq:freqVrDefinition}.

\subsection{Radiation pressure}
\label{sec:radPress}
\noindent
Radiation pressure is relevant only for H atoms because of a much lower spectral flux and the mass of He atoms being four-fold larger than that of H atoms. Therefore, from this point on we will be referring only to the quantities characteristic for this chemical element.
Each scattering event transfers momentum between the interacting photon and atom. A cumulative effect of many scatterings can be described as a continuous force acting on atoms up to a distance $\text{r}\sim 5000$ au from the Sun. Beyond this distance, the mean time between scattering events is comparable to the time needed for an atom to travel the length equivalent to its heliospheric distance and the continuous force approximation is no longer valid (see section 3.2.11 in \citet{hall:92a}). 
The continuous force caused by the momentum transfer, based on Equations \ref{eq:deltaP}, \ref{eq:gfactor}, can be expressed as:
\begin{equation}
\label{eq:prad}
\text{P}_\text{rad}=\text{g}\, \Delta \text{p} =\text{I}_\text{Lya}(\lambda_\text{H}) \frac{\pi \text{e}^2}{\text{m}_e \text{c}} \frac{\text{h}\lambda_\text{H}}{\text{c}} \text{f}_{\text{osc},\text{H}}.
\end{equation}

Radiation pressure acts in the opposite direction to the solar gravitational force $\text{F}_\text{g}=-\frac{\text{GM}_\sun\text{m}_\text{H}}{\text{r}^2}$. 
Since in the optically thin approximation the solar spectral flux drops with the square of solar distance, then the radiation pressure force is directly proportional to that of solar gravity at all distances. 
We define a dimensionless factor $\mu$ that is very commonly used in the literature. It is defined as a ratio between the force caused by the momentum transfer due to scattering events and gravitational force at a given distance. When it is equal to 1, there is no effective force acting on a moving atom. Since we have direct measurements of solar irradiance at the distance $r=r_\text{E}$, we will use it in the following considerations.
\begin{align}
\label{eq:radpress}
\nonumber
\mu&=\frac{\text{P}_\text{rad}}{|\text{F}_\text{g}|}\\ \nonumber
&= \text{I}_\text{Lya}(\lambda_\text{H}) \frac{\pi \text{e}^2}{\text{m}_e \text{c}}\frac{\text{h}\lambda_\text{H}}{\text{c}}\text{f}_{\text{osc},\text{H}} \frac{\text{r}_\text{E}^2}{\text{GM}_\sun\text{m}_\text{H}}\\
&=\frac{\text{I}_{lya}(\lambda_\text{H})}{\text{p}_\text{H}},\\
\label{eq:ph}
\text{p}_\text{H} & = \left[\frac{\pi\, \text{e}^2}{\text{m}_\text{e}\,c}\,\frac{\text{h}\lambda_\text{H}}{\text{c}} \text{f}_{\text{osc},\text{H}} \frac{\text{r}_\text{E}^2}{\text{G}\,\text{M}_\sun \text{m}_\text{H}}\right]^{-1} = 3.34467\times 10^{12} & \text{ ph s}^{-1}\,\text{cm}^{-2}\,\text{nm}^{-1}.
\end{align}

When the single-scattering optically-thin approximation is not used, then still the radiation pressure force at a given distance to the Sun can be expressed as a fraction $\mu$ of the solar gravity force at this distance, but it is no longer distance-independent.

\section{Available measurements of the solar He I 58.4 nm line}
\label{sec:appendix1}
\noindent
One of the first measurements of the 58.4 nm solar line was given by \citet{hall_hinteregger:70a} who reported the radiance in the He~I~58.4~nm line on 11th March 1967, 09:45 UT equal to $0.89 \cdot 10^9$~phot~cm$^{-2}$ s$^{-1}$ and the helium continuum level (measured at 50.4~nm) equal to $0.5 \cdot 10^9$~phot~cm$^{-2}$ s$^{-1}$. For 15th May 1967, the flux was reported larger by a factor of 1.6 at 58.43~nm. 

Another measurement of this line was done by \citet{doschek_etal:74a}, who obtained a geocorona-uncorrected FWHM$_\lambda$ equal to $0.014 \pm 0.0015$6~nm and reported no departures from the Gaussian shape.

\citet{cushman_etal:75a} based his analysis of the 58.4 nm solar line on a rocket underflight on August 30, 1973. The resolution was $\sim 0.002$~nm, and since the measurement was taken from within the geocorona, there was a 0.003~nm wide geocoronal absorption feature in the profile, for which the measurement was compensated. The appearance of the resulting profile is kappa-like, even though the authors do not discuss the differences between the assumed Doppler (Gaussian) shape and that actually measured. In addition, there was an active region in the field of view, which did not cover the entire solar disk. The FWHM$_\lambda$ of the helium line was found to be 0.01~nm for the active region and 0.008~nm for the quiet Sun. The measured spectral irradiance at the line center for the quiet Sun observation was found at $\sim 2\cdot 10^{10}$~phot (s \AA \, cm$^2)^{-1}$. The total intensity was reported at $1.3 \cdot 10^9\,\text{phot }(\rm{cm}^2\, s)^{-1} \pm 40$\%. The disk coverage was $\sim 3$\%. 

Radiance variations in the solar He~I~58.4 nm line from 5th March 1996 until 8th August 1999 were presented by \citet{schuhle_etal:00a}. The radiance seems to increase approximately linearly in the log scale from 0.4 to 0.5~W~m$^{-2}$~sr$^{-1}$ when measured on a portion of the disk corresponding to quiet Sun. These authors also show relative variations of the network and cell areas, i.e., the brightest and the weakest regions in the disk. While variations are by a factor of 2, these measurement feature an increase from solar minimum to solar maximum conditions similar to that for the quiet Sun.  

\citet{delZanna_andretta:06a} provide a series of EUV irradiance in the helium line between $\sim 1998$ and $\sim 2006$ observed by CDS/SOHO and obtain the absolute values between 1.5 and $2.0 \cdot 10^9$~phot~cm$^{-2}$~s$^{-1}$ (for comparison, the total irradiance in the Lyman-$\alpha$ line during the minimum of solar activity is about $3.5 \cdot 10^{11}$~phot~cm$^{-2}$~s$^{-1}$). \citet{delZanna_andretta:15a} present a continuation of the irradiance measurement from 1998 until 2015, spanning an interval $(1.0 - 2.0)\cdot 10^9$~phot~cm$^{-2}$~s$^{-1}$. 

We adopt the profile of the solar He~I~58.4 nm line as a Doppler (Gaussian) with a non-zero background, parametrized by a radial velocity of a scattering atom relative to the Sun $v_r$. The central wavelength of the transition responsible for the 58.4 nm ISN He backscatter glow is $\lambda_{\text{He I 58.4}} = 58.43350$~nm (CHIANTI database), which is equivalent to the frequency $\nu_{\text{He I 58.4}} = \nu_0 = 5.1304895\cdot 10^{15}$~Hz. The relation between the atom rest frame frequencies and radial speeds are given by Equation \ref{eq:freqVrDefinition}.

The evolution of the width of the solar He~I~58.4~nm line was presented by \citet{mcmullin_etal:04a} based on SUMER/SOHO observations carried out between 1996 and 2001, i.e., from a solar minimum to a maximum. The conclusion was that the Doppler width $v_{\mathrm{D}}$ was constant during the solar cycle and equal to $36.5 \pm 1.7$~\kms, where the uncertainty is the standard deviation of the measurement. The minimum and maximum 10\% of the measured widths were found to be at 27.7 and 49.5~\kms, respectively. The relation between the widths expressed by the wavelengths and the Doppler speeds used by \citet{mcmullin_etal:04a} is $v_{\mathrm{D}} = (\Delta \lambda/\lambda) c$, identical to that we use (Equation~\ref{eq:defineDopplerShift}). 

\citet{mcmullin_etal:04a} state that the widths $v_{\mathrm{D}}$ were fitted to a ``single Gaussian function and a linear background'', but they do not provide the formula for the profile. \citet{lallement_etal:04b} refers to the paper by \citet{mcmullin_etal:04a} and states that based on the data from this paper, the widths of this line are between 92 and 155~m{\AA}, with the preferred value at 118~m{\AA}. A function defined as $\exp\left[-\left(\frac{x}{2 \sigma} \right)^2 \right]$ takes the value of $\frac{1}{2}$  for $x = \pm 2 \sqrt{\ln 2}\,\sigma$, so $\mathrm{FWHM} = 4 \sqrt{\ln 2} \sigma$, which corresponds to $\mathrm{FWHM}_v = 0.118/584\, c \cong 60.6$~\kms. From this relation one obtains that $2 \sigma = \mathrm{FWHM}_v/(2 \sqrt{\ln 2}) \cong 36.4$~\kms, which we adopt in the model. 

\bibliographystyle{aasjournal}
\bibliography{helioGlowPaperI}  

\end{document}